\documentclass[%
reprint,
superscriptaddress,
 amsmath,amssymb,
prb,
]{revtex4-2}
\usepackage{amssymb}
\usepackage{amsmath}
\usepackage{bbm}
\usepackage{bm}
\usepackage{epsfig}
\usepackage{color}
\usepackage{multirow}
\usepackage{graphicx}
\usepackage{booktabs}
\usepackage{dsfont}
\usepackage{tikz}
\usepackage{appendix}
\usepackage{makecell}
\usepackage{subfigure}
\usepackage{braket}
\usepackage{xcolor}
\usepackage[colorlinks,linkcolor=blue,anchorcolor=blue,citecolor=blue,urlcolor=blue]{hyperref}
\begin{document}

\title{Quantum Avalanche Stability of Many-Body Localization with Power-Law Interactions}

\author{Longhui Shen}
\affiliation{Department of Physics, Wuhan University of Technology, Wuhan 430070, China}

\author{Bin Guo}
\email{binguo@whut.edu.cn}
\affiliation{Department of Physics, Wuhan University of Technology, Wuhan 430070, China}

\author{Zhaoyu Sun}
\affiliation{School of Electrical and Electronic Engineering, Wuhan Polytechnic University, Wuhan 430023, China}

\date{\today} 

\begin{abstract}
We investigate the stability of the many-body localized phase against quantum avalanche instabilities in a one-dimensional Heisenberg spin chain with long-range power-law interactions ($V\propto r^{-\alpha}$). By combining exact diagonalization of static properties with Lindblad master equation simulations of open-system dynamics, we systematically map the interplay between interaction range and disorder strength. Our finite-size scaling analysis of entanglement entropy identifies a critical interaction exponent $\alpha_c \approx 2$, which separates a fragile regime, characterized by an exponentially diverging critical disorder, from a robust short-range regime. To rigorously test the system's resistance to avalanches, we couple the boundary to an infinite-temperature bath and track the propagation of the thermalization front into the localized bulk. We find that the characteristic thermalization time follows a unified scaling law, $T_{r_{\text{th}}} \sim \exp[\kappa(\alpha) LW]$ (herein, $L$ is the system size, and $W$ is the disorder intensity), which diverges exponentially with the product of system size and disorder strength. This suppression enables the derivation of a quantitative stability criterion, $W_{\text{stab}}(\alpha)$, representing the minimum critical disorder strength required to maintain avalanche stability. Our results confirm that the MBL phase remains asymptotically stable in the thermodynamic limit when disorder exceeds an interaction-dependent threshold, bridging theoretical debates on long-range MBL and providing a roadmap for observing these dynamics in experimental platforms such as Rydberg atom arrays.
\end{abstract}


\maketitle

\section{Introduction}\label{sec1}

The thermalization of isolated quantum many-body systems represents a frontier challenge in contemporary physics~\cite{Nandkishore2015, Yao2014, Ganeshan2015, Schreiber2015, DAlessio2016, Rigol2008}. The universal mechanism underlying this process is articulated by the Eigenstate Thermalization Hypothesis (ETH)~\cite{Deutsch1991, Srednicki1994, Rigol2008}, which posits that in a generic ergodic system, the efficient exchange of energy and particles between subsystems leads local observables to relax into a thermal equilibrium state determined solely by global conservation laws. However, quantum phases that defy this paradigm, such as Many-Body Localization (MBL), are of profound significance for both condensed matter theory and quantum information technology~\cite{Bahri2015, Cao2025, Wang2018, DeTomasi2017, Abanin2019, Smith2016, Parameswaran2018, Lukin2019, Smith2019}. Emerging in systems characterized by strong quenched disorder and finite interactions, MBL provides a robust framework for protecting initial quantum information from thermalization by preventing the transport of energy and charge~\cite{Choi2016,Abanin2019}. MBL phases exhibit a suite of non-ergodic signatures, most notably the emergence of a complete set of quasi-local integrals of motion (LIOMs)~\cite{Imbrie2017, Huse2014, Abanin2019, Serbyn2013a} and a characteristic logarithmic growth of entanglement entropy over time~\cite{Serbyn2013, Serbyn2015}, as well as distinct signatures in quantum mutual information~\cite{DeTomasi2017}. This slow entanglement growth serves as a definitive hallmark of the system's resilience against thermalization.

While the MBL phase is widely regarded as stable in finite-sized systems, its integrity in the thermodynamic limit has recently been questioned by the quantum avalanche mechanism~\cite{DeRoeck2017, Zhang2025, Sels2022, Peacock2023, Szoldra2024}. This instability describes a scenario wherein a rare, locally ergodic, low-disorder region, frequently termed a "thermal bubble" or "ergodic inclusion", hybridizes with the surrounding localized bulk~\cite{Gopalakrishnan2016, Agarwal2015, Agarwal2017}. Such hybridization can trigger a runaway thermalization process that ultimately destabilizes the MBL phase. The dynamics of these avalanches have been extensively explored in both theoretical and experimental contexts, rendering the asymptotic stability of the MBL phase in large-scale systems a persistent and highly debated topic~\cite{Sierant2022, Bera2017, Panda2020}.

Compounding this debate is the unresolved role of long-range interactions~\cite{Burin2015a, Nandkishore2017, Nosov2019, Hauke2015, Bhakuni2021, Bhakuni2020, Thomson2020, Cheng2023}. Although the vast majority of MBL research has focused on short-range models, rapidly advancing experimental platforms in Atomic, Molecular, and Optical (AMO) physics—such as trapped-ion crystals~\cite{Smith2016, Zhang2017} and Rydberg atom ~\cite{Saffman2010, Bernien2017, Lukin2019, Rispoli2019} arrays, naturally realize Hamiltonians with power-law decaying interactions ($V \propto 1/r^\alpha$). Consequently, elucidating the interplay between long-range interactions and quantum avalanches is of paramount theoretical and experimental importance~\cite{Nandkishore2017, Cheng2023}. Specifically, it remains an open question whether the enhanced connectivity provided by power-law interactions accelerates avalanche propagation, thereby rendering the MBL phase more susceptible to instability.

In this work, we address these open questions by investigating the stability of MBL systems with long-range power-law interactions ($  V \propto r^{-\alpha}  $) against quantum avalanches~\cite{Scocco2024, Yousefjani2023, Burin2015a,Tikhonov2018}. We quantify long-range effects through open-system dynamics and derive a phase diagram that resolves the debate on asymptotic stability. Our primary innovation combines static and dynamic methods: exact diagonalization for equilibrium properties (entanglement entropy and level statistics) with Lindblad master equation simulations of open-system dynamics. This hybrid approach enables a comprehensive mapping of the interplay between interaction range $\alpha$ and disorder strength $W$. In the static sector, finite-size scaling of the normalized half-chain entanglement entropy reveals a critical interaction exponent $\alpha_c \approx 2$~\cite{Burin2015}, delineating a fragile long-range regime (where the critical disorder $W_c$ diverges exponentially) from a robust short-range-like regime.

To probe avalanche resistance directly, we introduce a controlled open-system setup by coupling the chain's leftmost site to an infinite-temperature bath, serving as a proxy for a large ergodic inclusion and seeding a localized avalanche source. Tracking the propagation of the thermalization front via the staggered magnetization imbalance $\overline{\mathcal{I}}(t)$ and its relative form $\mathcal{I}_r(t)$, we uncover a unified scaling law for the characteristic thermalization time: $T_{r_{\text{th}}}\sim \exp[\kappa(\alpha)LW]$, which diverges exponentially with the product of system size and disorder strength in the deep MBL regime.By probing the dynamics of the thermalization front, we quantify the expansion of the thermalized region. We demonstrate that deep within the MBL phase, thermalization progresses with a slow logarithmic growth over time, consistent with the analytical prediction of a "logarithmic light cone" in localized systems~\cite{Deng2017, Scocco2024}. This dynamical suppression establishes a lower bound on thermalization timescales and allows us to derive a quantitative stability criterion—the minimum disorder threshold $  W_{\text{stab}}(\alpha)  $ required for asymptotic avalanche stability in the thermodynamic limit. We present the resulting phase diagram of $W_{\text{stab}}(\alpha)  $, which highlights how short-range interactions enhance robustness while long-range connectivity promotes fragility, offering insights into MBL behavior in AMO quantum simulators.

The remainder of this paper is organized as follows. In Sec.~\ref{sec_model}, we introduce the one-dimensional Heisenberg model with power-law decaying interactions and detail the numerical methods used in our study, including exact diagonalization (ED) for static properties and the Lindblad master equation simulations of open-system dynamics. In Sec.~\ref{sec_isolated}, we analyze the isolated model at infinite temperature and construct the ETH--MBL phase diagram in the $(\alpha,W)$ plane using entanglement-entropy and level-statistics diagnostics with finite-size scaling. In Sec.~\ref{sec_avalanche}, we present our main results on avalanche stability by coupling one end of the chain to an infinite-temperature bath, tracking the bath-induced thermalization dynamics, and quantifying how the relevant thermalization timescale scales with system size, disorder strength, and interaction range. Finally, Sec.~\ref{sec_conclusion} summarizes our findings and discusses open questions and future directions.

\section{Model and formulas}\label{sec_model}

We study the dynamics of a one-dimensional spin-1/2 $XXZ$ Heisenberg chain with power-law decaying interactions in the presence of a random magnetic field. The Hamiltonian is given by~\cite{Yousefjani2023, SafaviNaini2019, Deng2020}
\begin{align}
	\hat{H} = & -\sum_{i \neq j}^{L} \frac{J_x}{|i-j|^\alpha} \left\{ J_x (\hat{S}_i^x \hat{S}_j^x + \hat{S}_i^y \hat{S}_j^y) + J_z\hat{S}_i^z \hat{S}_j^z \right\} \nonumber\\
	& + \sum_{i=1}^{L} h_i \hat{S}_i^z,
\end{align}
where $L$ denotes the system size, and $\hat{S}^\mu_j = \hat{\sigma}^\mu_j/2$ are the local spin operators ($\mu \in \{x,y,z\}$). We set the energy scales $J_x = J_z = 1$. The local magnetic fields $h_i$ are drawn uniformly from the interval $[-W, W]$. The exponent $\alpha$ governs the power-law decay of the interaction strength with distance, thereby determining the effective range of the spin couplings. The model conserves the total magnetization in the $z$-direction, $\hat{S}^z_{\text{tot}} = \sum_{j=1}^L \hat{S}^z_j$. 

Prior to investigating the open system dynamics, we characterize the static properties of the model in the infinite-temperature limit using Exact Diagonalization. These calculations are performed in the $\hat{S}^z_{\text{tot}} = 0$ symmetry sector using the QuSpin package~\cite{Weinberg2019}. To analyze the phase transition, we calculate the half-chain entanglement entropy averaging over the middle $100$ eigenstates of the spectrum. The disorder averaging is performed over $N_r$ realizations, ranging from $N_r = 5000$ for small systems to $N_r = 200$ for the largest system size considered, $L=16$.

Previous studies~\cite{SafaviNaini2019,Schiffer2019,Li2016,Deng2020} have systematically investigated how the ETH--MBL transition in isolated systems depends on the interaction exponent $\alpha$, and reported strong finite-size trends in the critical disorder strength. While these results capture the qualitative crossover from ergodic to localized behavior, their direct extrapolation to the thermodynamic limit remains challenging, especially in the presence of long-range couplings where rare-region effects and resonances are strongly enhanced. As a consequence, the stability of the MBL phase under long-range interactions remains an open question. In this work, we address this issue from a dynamical perspective by coupling one end of the system to a thermal bath and directly probing avalanche-induced delocalization.

To simulate the effect of a quantum avalanche or an infinite ergodic region, we couple the leftmost site ($i=1$) of the chain to a thermal bath. The time evolution of the system's density matrix $\hat{\rho}(t)$ is governed by the Lindblad master equation~\cite{Sels2022,Tu2023, Scocco2024}
\begin{align}
	\frac{d}{dt}\hat{\rho}(t) = -i[\hat{H}, \hat{\rho}(t)] + \mathcal{D}[\hat{\rho}(t)],
\end{align}
where we set $\hbar=1$. To induce infinite-temperature ($T=\infty$) correlations while preserving the particle number (magnetization) conservation symmetry, we employ a dephasing-like dissipator acting on the first spin. The dissipation term is explicitly given by the double commutator form
\begin{align}
	\mathcal{D}[\hat{\rho}(t)] = -\gamma^2 [\hat{S}_1^z, [\hat{S}_1^z, \hat{\rho}(t)]].
\end{align}
here, $\gamma$ represents the coupling strength to the bath. This corresponds to a single collapse operator $\hat{C} = \sqrt{2}\gamma \hat{S}_z^1$. As noted in Ref.~\cite{Scocco2024}, such particle-conserving noise does not alter the fundamental logarithmic slope of transport in MBL systems. We restrict our calculations to a specific $\hat{S}^z_{\text{tot}}$ subspace; within this subspace, the identity matrix serves as the unique steady state of the Lindblad dynamics, implying that the bath effectively drives the system toward infinite-temperature thermalization.

We study the non-equilibrium dynamics by initializing the system in the high-energy Néel product state, $|\psi(0)\rangle = |\uparrow\downarrow\uparrow\downarrow\cdots\rangle$. To numerically solve the Lindblad master equation, we employ the quantum trajectory method (also known as the Monte Carlo Wave Function algorithm) as implemented in the QuTiP software package~\cite{Lambert2026}. For the open system calculations, the coupling strength to the thermal bath is fixed at $\gamma=1$. The system's evolution is tracked over a logarithmic time grid spanning $t \in [0.1, 1000]$. Statistical convergence is ensured by averaging over $10–50$ independent stochastic trajectories for each disorder realization across the various system sizes considered. The results are further averaged over a disorder ensemble ranging from $500$ realizations for the smallest system size ($L=8$) to $100$ realizations for the largest system size ($L=16$).

To rigorously isolate the thermalization effects induced by the boundary bath, we benchmark the dynamics of the open system ($\gamma=1$) against the corresponding isolated system ($\gamma=0$). Crucially, this comparison is performed using the identical set of disorder realizations to minimize statistical variance and accurately quantify the propagation of the thermal front.

\section{Interaction-Range Dependence of the MBL Transition}\label{sec_isolated}

\begin{figure}[t!]
\centering
\includegraphics[width=0.45\textwidth,keepaspectratio]{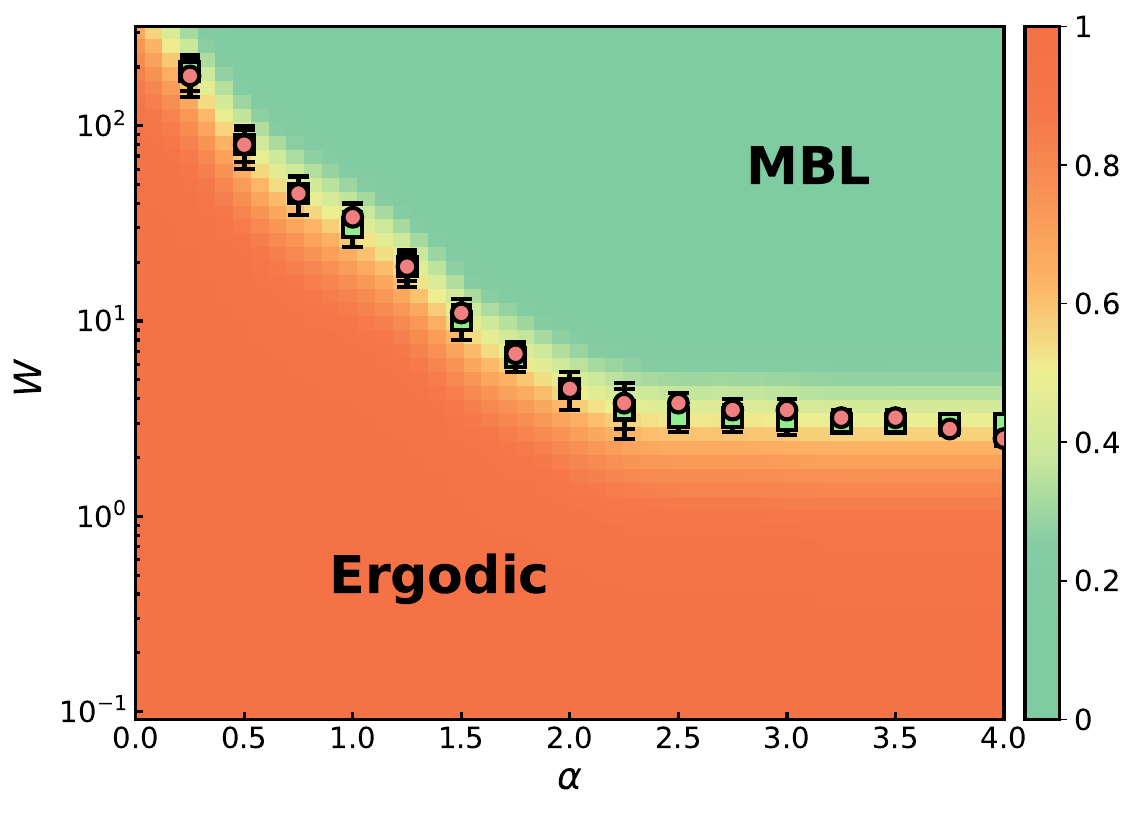}
\caption{Phase diagram of the one-dimensional Heisenberg chain in the $(\alpha, W)$ plane with power-law interactions. The colormap represents the normalized half-chain entanglement entropy $\langle S \rangle  /S_T$ and the normalized level statistics ratio $r$. The data reveal two distinct phases: an ergodic phase (orange/red regions), characterized by volume-law entanglement ($\langle S \rangle \approx S_T$), and a many-body localized (MBL) phase (green regions), characterized by area-law entanglement ($\langle S \rangle \ll S_T$). Circular markers with error bars indicate the estimated critical disorder strength $W_c(\alpha)$, which delineates the ETH--MBL transition boundary. The phase boundary indicates that as the interaction range increases (decreasing $\alpha$), the MBL phase requires substantially stronger disorder to remain stable.}
\label{Fig1}
\end{figure}

To systematically explore the interplay between interaction range and disorder, we construct the phase diagram in the $(\alpha, W)$ plane, as shown in Fig.~\ref{Fig1}. The background color encodes the degree of ergodicity, quantified by the normalized entanglement entropy and level statistics ratio, which yield consistent boundary demarcations. A clear boundary separates the ergodic region (orange/red, high entropy) from the localized region (green, low entropy). We observe a monotonic relationship between the critical disorder $W_c$ and the interaction exponent $\alpha$. This trend physically reflects the enhanced delocalization tendency driven by long-range coupling: as $\alpha$ decreases, the effective connectivity of the spin chain increases, thereby requiring a stronger random potential to suppress quantum transport.

Based on the transition boundary extracted in Fig.~\ref{Fig1}, the interaction-range dependence of the critical disorder reveals $\alpha \approx 2$ as a pronounced crossover point. In the strong long-range regime ($\alpha<2$), achieving localization becomes increasingly difficult; the critical disorder $W_c$ rises drastically, suggesting a divergence toward infinity as the interaction range increases. This divergence is consistent with the theoretical prediction by Burin~\cite{Burin2015}, who argued that for $\alpha < 2d$ (where $d=1$), the proliferation of resonant long-range excitations inevitably destabilizes the MBL phase~\cite{Burin2015a, Modak2020}. Consequently, for very small $\alpha$, $W_c$ can exceed $W \sim 10^2$, indicating that the system is highly susceptible to delocalization unless the disorder is exceptionally strong. In contrast, in the short-range limit ($\alpha \gtrsim 2$), the phase boundary tends toward saturation, with the crossing points of the half-chain entanglement entropy eventually reaching a constant value of $W_c \approx 3.3$. This asymptotic behavior is similar to the established critical disorder of the standard nearest-neighbor Heisenberg model ($W_c \approx 3.7$)~\cite{Yousefjani2023a}, confirming that for $\alpha \gtrsim 3$, the long-range tail of the interaction becomes irrelevant to the localization physics.

\begin{figure}[t!]
\centering
\includegraphics[width=0.4\textwidth,keepaspectratio]{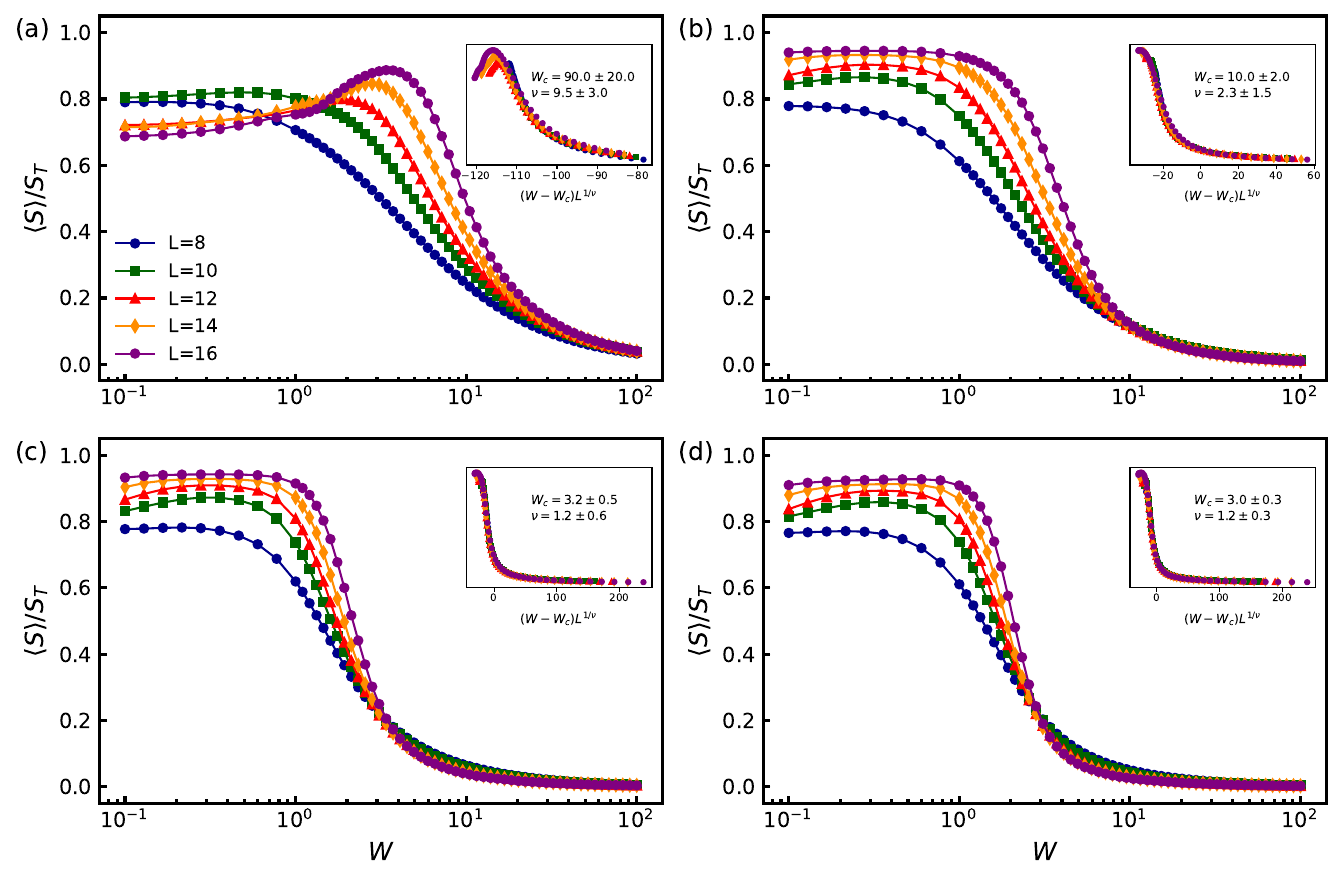}
\caption{Finite-size scaling analysis of the normalized half-chain entanglement entropy $\langle S \rangle / S_T$ as a function of disorder strength $W$ for four representative interaction exponents: (a) $\alpha = 0.5$, (b) $\alpha = 1.5$, (c) $\alpha = 2.5$, and (d) $\alpha = 3.5$. The main panels show the entropy ratio for system sizes $L=8, 10, 12, 14, 16$ (represented by different colors and symbols). The crossing of the curves signifies the transition from the volume-law ergodic phase to the area-law MBL phase. The insets display the data collapse obtained using the scaling ansatz $f[(W - W_c)L^{1/\nu}]$, which allows for the extraction of the critical disorder $W_c$ and the critical exponent $\nu$ listed in each panel~\cite{Yousefjani2023a}. The excellent collapse confirms the validity of the estimated critical points used in the phase diagram.}
\label{Fig2}
\end{figure}

To quantitatively extract the critical disorder strength and characterize the transition, we perform a finite-size scaling analysis of the normalized entanglement entropy. Figure~\ref{Fig2} presents the behavior of $\langle S \rangle / S_T$ as a function of disorder $W$ for distinct interaction regimes, where $S_T = \frac{1}{2} \left[ L \log(2) - 1 \right]$ is the Page value~\cite{Page1993, Khemani2017}. In all cases, we observe a clear crossover from a thermal phase ($\langle S \rangle / S_T \approx 1$) at weak disorder to a localized phase ($\langle S \rangle / S_T \to 0$) at strong disorder. To pinpoint the critical disorder $W_c$, we posit that in the vicinity of the MBL transition, the entropy follows the single-parameter scaling ansatz~\cite{Khemani2017, Sierant2025, Yousefjani2023a}
\begin{align}
	\langle S \rangle / S_T = \tilde{f}\left[ \frac{L}{\xi(W)} \right] = f\left((W-W_c)L^{1/\nu}\right),
\end{align}
where $\xi(W) \propto |W - W_c|^{-\nu}$ represents the diverging correlation length near the transition, and $\nu$ denotes the critical exponent.

As demonstrated in the insets of Fig.~\ref{Fig2}, rescaling the disorder axis by $(W - W_c)L^{1/\nu}$ yields an excellent data collapse for different system sizes across the various interaction regimes. This analysis quantifies the dramatic shift in stability discussed in the phase diagram: for the long-range case of $\alpha = 0.5$, the system remains ergodic up to extremely high disorder values ($W_c \approx 90$), whereas for short-range interactions ($\alpha = 3.5$), the critical point stabilizes at $W_c \approx 3.0$. The successful scaling collapse serves as compelling evidence for the existence of a sharp MBL transition. This verification is particularly crucial for the intermediate case of $\alpha=2.5$, where relying solely on the visual inspection of crossing points in the raw data can be inconclusive due to finite-size drifts. However, we note a significant feature regarding the fitting procedure: while the data collapse for the long-range regimes ($\alpha=0.5$ and $1.5$) is visually robust, the extracted fitting parameters, specifically $W_c$ and $\nu$, carry relatively large uncertainties. These larger errors arise from the distinct and fragile nature of the transition in the presence of strong long-range interactions, a feature we will discuss in greater detail below.

\begin{figure}[t!]
\centering
\includegraphics[width=0.45\textwidth,keepaspectratio]{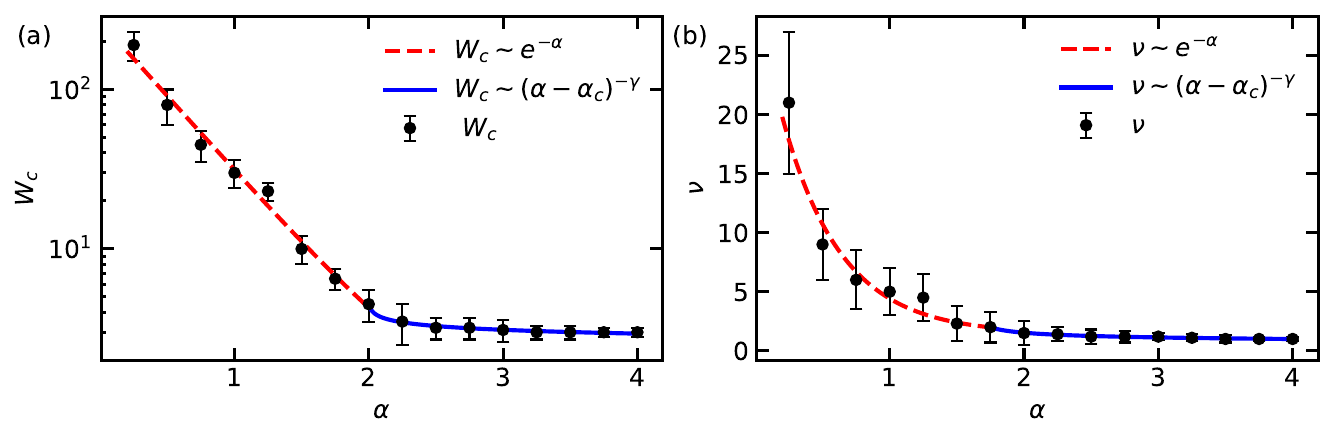}
\caption{Finite-size scaling results for the MBL transition parameters as a function of the interaction exponent $\alpha$. (a) The critical disorder strength $W_c$. (b) The critical exponent $\nu$. In both panels, black circles with error bars represent values extracted from the scaling analysis of the normalized entanglement entropy $\langle S \rangle/S_T$. The data reveals two distinct regimes separated by a critical interaction exponent $\alpha_c \approx 2$. The blue solid lines denote power-law fits [Eq.~(\ref{Eq.5})] in the short-range regime ($\alpha > 2$), illustrating the divergence near $\alpha_c$. The red dashed lines indicate exponential fits in the long-range regime ($\alpha < 2$), highlighting the rapid growth of both the critical disorder and the critical exponent as interactions become longer-ranged.}
\label{Fig3}
\end{figure}

We extend the finite-size scaling analysis of $\langle S \rangle / S_T$ to a broad range of interaction exponents $\alpha \in [0.5,4]$ to explicitly map out the ETH--MBL transition boundary $W_c(\alpha)$. The resulting critical disorder strength $W_c$ and critical exponent $\nu$ are summarized in Fig.~\ref{Fig3}, respectively. When $\alpha \leq 2.0$, we observe a marked divergence in both critical parameters, accompanied by a sharp increase in statistical uncertainty. This behavior strongly suggests the existence of a critical interaction exponent, $\alpha_c$, below which the MBL phase becomes unstable or requires unphysically large disorder levels. Based on this trend, we categorize the physics into two regimes: a short-range-like regime for $\alpha > 2$ and a long-range exponential growth regime for $\alpha < 2$.

To quantitatively determine the value of $\alpha_c$, we adopt the method outlined in Ref.~\cite{Schiffer2019} and model the divergence of the critical parameters using a power-law ansatz
\begin{align}\label{Eq.5}
	\eta(\alpha) = A_{\eta} (\alpha - \alpha_{c,\eta})^{-\gamma_{\eta}},
\end{align}
where $\eta$ represents either $W_c$ or $\nu$. Due to the limited system sizes accessible in ED, the true divergence near $\alpha_c$ is suppressed. To mitigate this finite-size effect, we introduce a lower bound cutoff $\alpha_f$ and perform the fit only for data satisfying $\alpha > \alpha_f$. The selection of this cutoff is critical for accurately capturing the asymptotic behavior near the singularity while avoiding the crossover regime where the power-law ansatz breaks down. To ensure the robustness of our results, we performed a systematic stability analysis by monitoring the relative fitting errors of the parameters ($A, \alpha_c, \gamma$) as a function of the cutoff $\alpha_f$. The optimal $\alpha_f$ was identified as the point where these errors were minimized and the extracted parameters exhibited stability. Detailed plots of this error analysis and the stability regions are provided in Appendix~\ref{sec:appendix_stability}.

For the critical disorder $W_c$ shown in Fig.~\ref{Fig3}(a), error analysis yields an optimal cutoff of $\alpha_f \approx 2$. The resulting best-fit parameters are $A_W = 3.11 \pm 0.02$, $\gamma_W = 0.08 \pm 0.01$, and a critical interaction exponent $\alpha_{c,W} = 1.990 \pm 0.006$. Similarly, for the critical exponent $\nu$ in Fig.~\ref{Fig3}(b), we find an optimal cutoff range $\alpha_f \in [1.25, 2.25]$, with a global minimum at $\alpha_f = 1.75$. The corresponding best-fit parameters are $A_{\nu} = 1.2117 \pm 0.0239$, $\gamma_{\nu} = 0.23 \pm 0.03$, and $\alpha_{c,\nu} = 1.64 \pm 0.04$. Although the scaling exponents $\gamma$ differ, the estimates for $\alpha_c$ are broadly consistent, supporting the conclusion that $\alpha_c \approx 2$, which aligns with previous literature~\cite{Burin2015, Modak2020}. It should be noted that the errors reported here reflect only the precision of the numerical fitting, and do not fully account for the inherent uncertainty of the scaling behavior near the divergence point.

In the regime of strong long-range interactions ($\alpha < 2$), the power-law description breaks down. We find that the behavior of $W_c$ is well-captured by an exponential function (red dashed line in Fig.~\ref{Fig3}(a))
\begin{align}
	W_{c}(\alpha) = 266.47 \cdot e^{-2.16\alpha} + 0.76.
\end{align}

This exponential dependence confirms that for $\alpha < 2$, the delocalizing tendency of long-range interactions dominates, causing the critical disorder to scale drastically. Remarkably, as shown in Fig.~\ref{Fig3}(b), the critical exponent $\nu$ displays a similar exponential divergence in this regime, further underscoring the distinct universality class of the transition at small $\alpha$

\begin{align}
	\nu(\alpha) = 28.94  \cdot e^{-2.28\alpha}+1.44.
\end{align}

Finally, we address the implications of our results for the universality class of the transition, focusing on the critical exponent $\nu(\alpha)$. For $\alpha > 2.25$, our extracted values yield $\nu \approx 1$. While this is typical for ED studies of short-range MBL models, it violates the strict Harris/CCFS/CLO bound ($\nu \ge 2$) required for stability in one-dimensional disordered systems~\cite{Luitz2015, Khemani2017, Sierant2023}. We attribute this discrepancy to finite-size effects, as the full extent of quenched randomness is not realized at the system sizes ($L \le 16$) accessible via exact diagonalization. However, a key observation is the smooth crossover behavior: as $\alpha$ decreases towards $\alpha_c$, $\nu(\alpha)$ gradually increases, eventually satisfying and exceeding the Harris bound. This continuous variation suggests that the MBL transition in systems with power-law interactions likely belongs to the same universality class as the short-range model, appearing distinct only due to severe finite-size drifts in the short-range limit~\cite{Burin2015a, Schiffer2019, Nag2019}. Although our current finite-size scaling provides robust evidence for the phase boundaries and this crossover, definitively resolving the critical exponents in the thermodynamic limit remains a challenge. We anticipate that future studies utilizing tensor network algorithms~\cite{Modak2020, Wahl2017, Sierant2022}, which can access significantly larger system sizes, will be instrumental in validating these scaling behaviors and strictly verifying the Harris bound.

\section{Robustness of the MBL Phase against Avalanche-Induced Thermalization}\label{sec_avalanche}

\begin{figure*}[t!]
\centering
\includegraphics[width=1\textwidth,keepaspectratio]{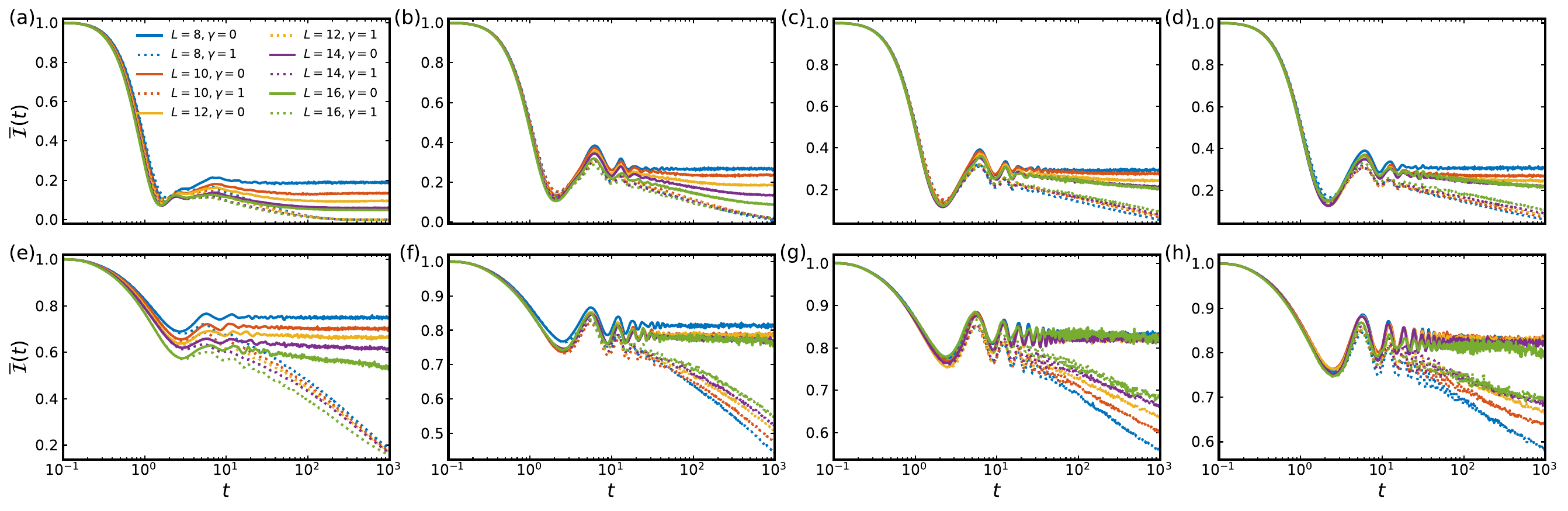}
\caption{Dynamics of the magnetization imbalance $\mathcal{I}(t)$ for system sizes $L \in \{8, 10, 12, 14,16\}$. The columns correspond to interaction exponents $\alpha = 0.5, 1.5, 2.5, 3.5$ from left to right. (a)-(d) Weak disorder regime ($W=2$). (e)-(h) Strong disorder regime ($W=15$). In all panels, solid lines represent the isolated system ($\gamma=0$), while dotted lines represent the open system coupled to a thermal bath ($\gamma=1$). In the ergodic regime (top row), the system thermalizes rapidly regardless of $\alpha$. In the strong disorder regime (bottom row), a sharp contrast emerges: for isolated systems, stable MBL plateaus appear for large $\alpha$. When coupled to the bath, the decay rate exhibits distinct size dependence; for $\alpha < 1.5$, the curves collapse, indicating size-independent instability, whereas for $\alpha > 1.5$, the decay significantly slows down with increasing $L$, signaling robustness against avalanches.}
\label{Fig4}
\end{figure*}

In this section, we investigate the stability of the MBL phase against avalanche instabilities in the presence of long-range interactions. Guided by the static results from the preceding section, we identify relevant parameter regimes and couple the system boundary to a thermal bath to analyze the propagation dynamics of the thermalization front. The characteristically slow, logarithmic expansion of this front allows us to establish a lower bound for the thermalization timescale and map the stability boundary in the $W –\alpha$ parameter space. We find that within this localized regime, the thermalization timescale diverges exponentially with system size, thereby demonstrating the robust resistance of the MBL phase to avalanche-induced thermalization.

To quantify the real-time dynamics, we employ the magnetization imbalance $\overline{\mathcal{I}}(t)$, defined as the normalized difference between the spin expectation values on even and odd lattice sites~\cite{Doggen2019, Scocco2024, SafaviNaini2019, Doggen2018}
\begin{align}
	\overline{\mathcal{I}}(t) = \frac{1}{L} \sum_{j=1}^{L} (-1)^j \langle \psi(t) | \hat{S}_j^z | \psi(t) \rangle.
\end{align}

This global observable originates from local quantities and is experimentally accessible. The normalization ensures that $\overline{\mathcal{I}}(0) = 1$. The long-time asymptotic value of the magnetization imbalance $\overline{\mathcal{I}}(t)$ serves as an effective dynamical order parameter for the MBL phase; in the ergodic phase (small disorder $W$), $\overline{\mathcal{I}}(t)$ typically follows a power-law decay $\overline{\mathcal{I}}(t) \propto t^{-\beta}$, whereas in the finite-size localized phase (large $W$), numerical simulations indicate that it saturates to a non-zero finite value. Our objective is to analyze how the imbalance in the open system, $\overline{\mathcal{I}}_{\gamma=1}(t)$, deviates from that of the isolated system, $\overline{\mathcal{I}}_{\gamma=0}(t)$, and to investigate its scaling behavior with respect to the system size $L$, disorder strength $W$, and interaction range $\alpha$.

Figure~\ref{Fig4} presents typical examples of the temporal evolution of $\overline{\mathcal{I}}(t)$ for system sizes $L=8$ to $16$ and interaction exponents $\alpha \in \{0.5, 1.5, 2.5, 3.5\}$. Panels (a)-(d) of Fig.~\ref{Fig4} correspond to the weak disorder regime ($W=2$), while panels (e)-(h) depict the strong disorder regime ($W=15$). A general trend is observed: upon coupling to the thermal bath ($\gamma \neq 0$), the dynamics begin to deviate from the isolated trajectory $\overline{\mathcal{I}}_{\gamma=0}(t)$ around $t \sim 1$ and starts decaying toward zero. Notably, increasing the value of $\alpha$ significantly slows down this decay; a larger $\alpha$ leads to a prolonged interval before $\overline{\mathcal{I}}_{\gamma=1}(t)$ approaches zero.

Based on our parameter classification in the previous section, Figs.~\ref{Fig4}(a)–\ref{Fig4}(e) correspond to the ETH phase, and Figs.~\ref{Fig4}(f)–\ref{Fig4}(h) correspond to the MBL phase, with Figs.~\ref{Fig4}(c),~\ref{Fig4}(d), and~\ref{Fig4}(f) located near the phase transition boundary. In the weakly disordered regime ($W=2$, Figs.~\ref{Fig4}(a)–\ref{Fig4}(d)), for all $\alpha$ values in the isolation limit ($\gamma=0$), the magnetization imbalance $\overline{\mathcal{I}}_{\gamma=0}(t)$ remains low within the simulation window ($t < 10^3$). In the strongly disordered regime ($W=15$, Figs.~\ref{Fig4}(e)–\ref{Fig4}(h)), the uncoupled system typically exhibits a slowly decaying plateau with a high $\overline{\mathcal{I}}_{\gamma=0}(t)$ value. In Figs.~\ref{Fig4}(f)–\ref{Fig4}(h), this slow decline is minimal. For Figs.~\ref{Fig4}(g) and~\ref{Fig4}(h), the data for all $L$ converge to a single plateau. In Fig.~\ref{Fig4}(f) ($\alpha=1.5$), the isolated system approaches a plateau but does not fully converge. The stability of this plateau depends significantly on $\alpha$, becoming more robust as $\alpha$ increases. For the $\overline{\mathcal{I}}_{\gamma=0}(t)$ results in Figs.~\ref{Fig4}(a)–\ref{Fig4}(e), the system exhibits pronounced system-size dependence. Specifically, larger $L$ results in lower plateau values, particularly for small $\alpha$. This suggests that for small $\alpha$, the plateau value vanishes in the thermodynamic limit ($L \to \infty$).

Upon introducing thermal coupling (dashed lines), $\overline{\mathcal{I}}_{\gamma=1}(t)$ in Figs.~\ref{Fig4}(a)–\ref{Fig4}(e) rapidly decays to zero, especially for lower $\alpha$ values, indicating that long-range interactions facilitate rapid thermalization propagation. Even for higher $\alpha$ values, the decay rate slows slightly, but the imbalance $\overline{\mathcal{I}}(t)$ eventually vanishes after sufficient time, confirming that under the ETH mechanism, the thermalization process is rapid regardless of the interaction range. We observe that the trajectories of $\overline{\mathcal{I}}_{\gamma=1}(t)$ in Figs.~\ref{Fig4}(a),~\ref{Fig4}(b), and~\ref{Fig4}(e) collapse onto a single curve across all system sizes, indicating that avalanche propagation is uniform and independent of system size deep within the ETH region. However, in Figs.~\ref{Fig4}(c) and~\ref{Fig4}(d), while $\overline{\mathcal{I}}_{\gamma=1}(t)$ tends to collapse into a broader curve, it shows signs of curve separation, indicating that the thermalization rate is still influenced by system size during the transition from ETH to MBL. In contrast, Figs.~\ref{Fig4}(f),~\ref{Fig4}(g), and~\ref{Fig4}(h) in the MBL and asymptotic transition regions exhibit strong finite-size dependence. Under thermal coupling, the decay curves $\overline{\mathcal{I}}_{\gamma=1}(t)$ separate from one another; as $L$ increases, the rate of approach to zero slows down significantly. This trend is most pronounced in Figs.~\ref{Fig4}(g) and~\ref{Fig4}(h) (larger $\alpha$), where we observe a clear separation of timescales: the thermalization rate decreases significantly with increasing $L$. Therefore, we can conclude that although thermal coupling ultimately leads to thermalization, the physical processes under different mechanisms are fundamentally distinct. Deep within the ETH phase, avalanche propagation is rapid and size-independent. Conversely, deep within the MBL phase, the thermalization process is drastically inhibited by a dynamical suppression that scales with system size.

\begin{figure*}[t!]
\centering
\includegraphics[width=1\textwidth,keepaspectratio]{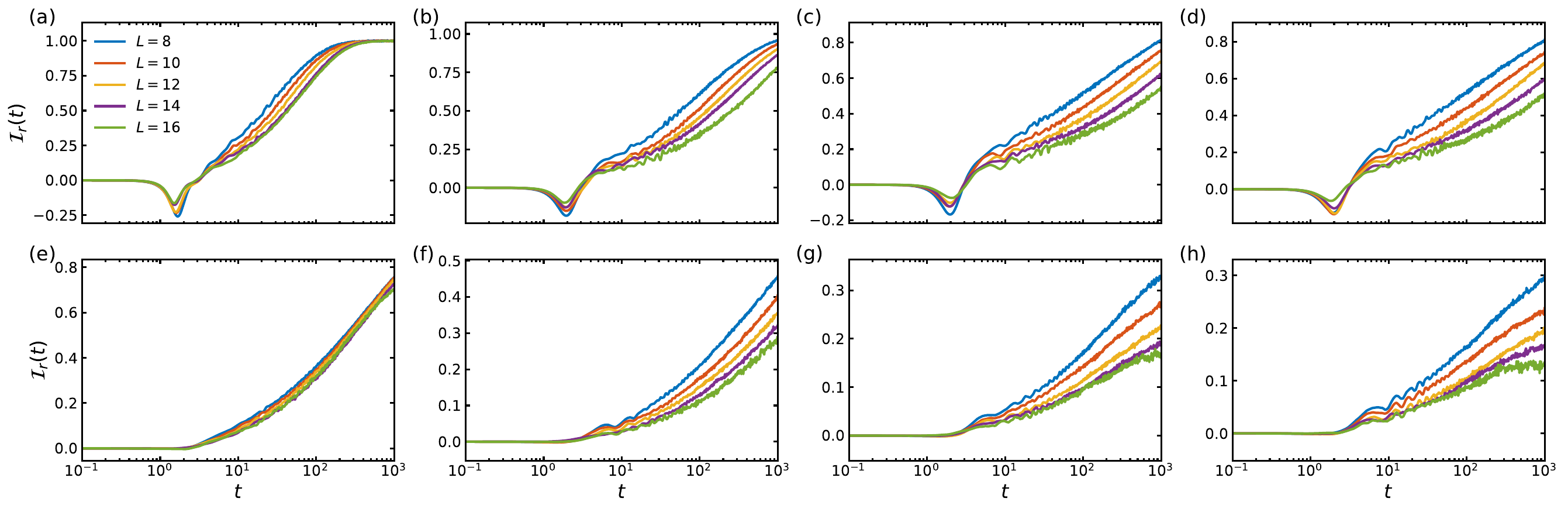}
\caption{Finite-size scaling of the relative magnetization imbalance $\mathcal{I}_r(t)$ for system sizes $L \in \{8, 10, 12, 14, 16\}$. The panels are organized by disorder strength and interaction range: (a)–(d) the weak disorder regime ($W=2$), with $\alpha \in \{0.5, 1.5, 2.5, 3.5\}$ from left to right; (e)–(h) the strong disorder regime ($W=15$), with corresponding $\alpha$ values.}
\label{Fig5}
\end{figure*}

To quantitatively characterize the propagation speed of the quantum avalanche and facilitate a finite-size scaling analysis, we introduce the relative magnetization imbalance $\mathcal{I}_r(t)$, as defined in Ref.~\cite{Scocco2024}
\begin{equation}
\mathcal{I}_r(t) = \frac{\overline{I}_0(t) - \overline{I}_\gamma(t)}{\overline{I}_0(t)}.
\end{equation}

This metric effectively isolates the thermalization induced solely by the bath coupling. Physically, $\mathcal{I}_r(t)$ increases monotonically from an initial value of $\mathcal{I}_r(0) = 0$ to an asymptotic value of $\mathcal{I}_r(t \to \infty) = 1$ as the system fully thermalizes. Consequently, the interaction exponent $\alpha$ directly dictates the rate of this growth: a larger $\alpha$ (shorter-range interaction) corresponds to a slower decay of the raw imbalance $\overline{I}_\gamma(t)$, and thus a significantly longer time is required for $\mathcal{I}_r(t)$ to approach saturation.

To quantitatively evaluate the avalanche stability of the system, we analyze the finite-size dependence of the relative imbalance $\mathcal{I}_r(t)$ across various interaction regimes. Figure~\ref{Fig5} illustrates the temporal evolution of $\mathcal{I}_r(t)$ for system sizes $L \in \{8, 10, 12, 14, 16\}$, comparing the dynamical response of the ergodic phase, the localized phase, and the intermediate transition regions. The parameter configurations are consistent with those in Fig.~\ref{Fig4}, with interaction exponents $\alpha = [0.5, 1.5, 2.5, 3.5]$ arranged from left to right. Figures ~\ref{Fig5}(a)–\ref{Fig5}(d) represent the weakly disordered regime ($W=2$), while Figs.~\ref{Fig5}(e)–\ref{Fig5}(h) represent the strongly disordered regime ($W=15$). Based on our previous classification, Figs.~\ref{Fig5}(a)–\ref{Fig5}(e) reside within the ETH phase, Figs.~\ref{Fig5}(f)–\ref{Fig5}(h) characterize the MBL phase, and Figs.~\ref{Fig5}(c),~\ref{Fig5}(d), and~\ref{Fig5}(f) delineate the ETH–MBL crossover.

In the ETH phase [Figs.~\ref{Fig5}(a),~\ref{Fig5}(b), and~\ref{Fig5}(e)], the dynamics are dominated by rapid thermalization. Regardless of the interaction exponent $\alpha$, the curves for different system sizes exhibit nearly identical growth rates and eventually saturate at $\mathcal{I}_r = 1$. Notably, in Fig.~\ref{Fig5}(e), the $\mathcal{I}_r(t)$ trajectories for all system sizes collapse onto a single curve. This size-independent behavior implies that the avalanche propagation rate is constant and the thermalization timescale is insensitive to the system size. Such a lack of significant finite-size dependence confirms that, in the ergodic phase, bath-induced thermalization is a global bulk phenomenon where the avalanche effectively thermalizes the system at a rate independent of its length.

Conversely, in the MBL phase [Figs.~\ref{Fig5}(g) and~\ref{Fig5}(h)], the dynamics exhibit a characteristic 'fan-shaped' separation as $\alpha$ increases (i.e., as interactions become shorter-ranged). We observe that the thermalization process is monotonically suppressed with increasing system size; specifically, larger systems (e.g., $L=16$) show significantly slower growth in $\mathcal{I}_r(t)$. This systematic curve separation indicates that in the short-range regime, the localized bulk exerts a robust resistance to avalanche propagation, causing the characteristic thermalization time to diverge as the system size increases. This size-dependent slowdown provides strong numerical evidence that the MBL phase remains stable against quantum avalanches in the thermodynamic limit, provided the interaction range is sufficiently short.Finally, the intermediate transition regions [Figs.~\ref{Fig5}(c),~\ref{Fig5}(d), and~\ref{Fig5}(f)] display hybrid characteristics. In Figs.~\ref{Fig5}(c) and~\ref{Fig5}(d), while the growth of $\mathcal{I}_r(t)$ remains rapid, the trajectories begin to separate with increasing $L$. This indicates that even at lower disorder strengths, the thermalization front begins to experience resistance as a function of system size, marking the onset of the crossover from ergodic to localized dynamics.

\begin{figure}[t!]
\centering
\includegraphics[width=0.45\textwidth,keepaspectratio]{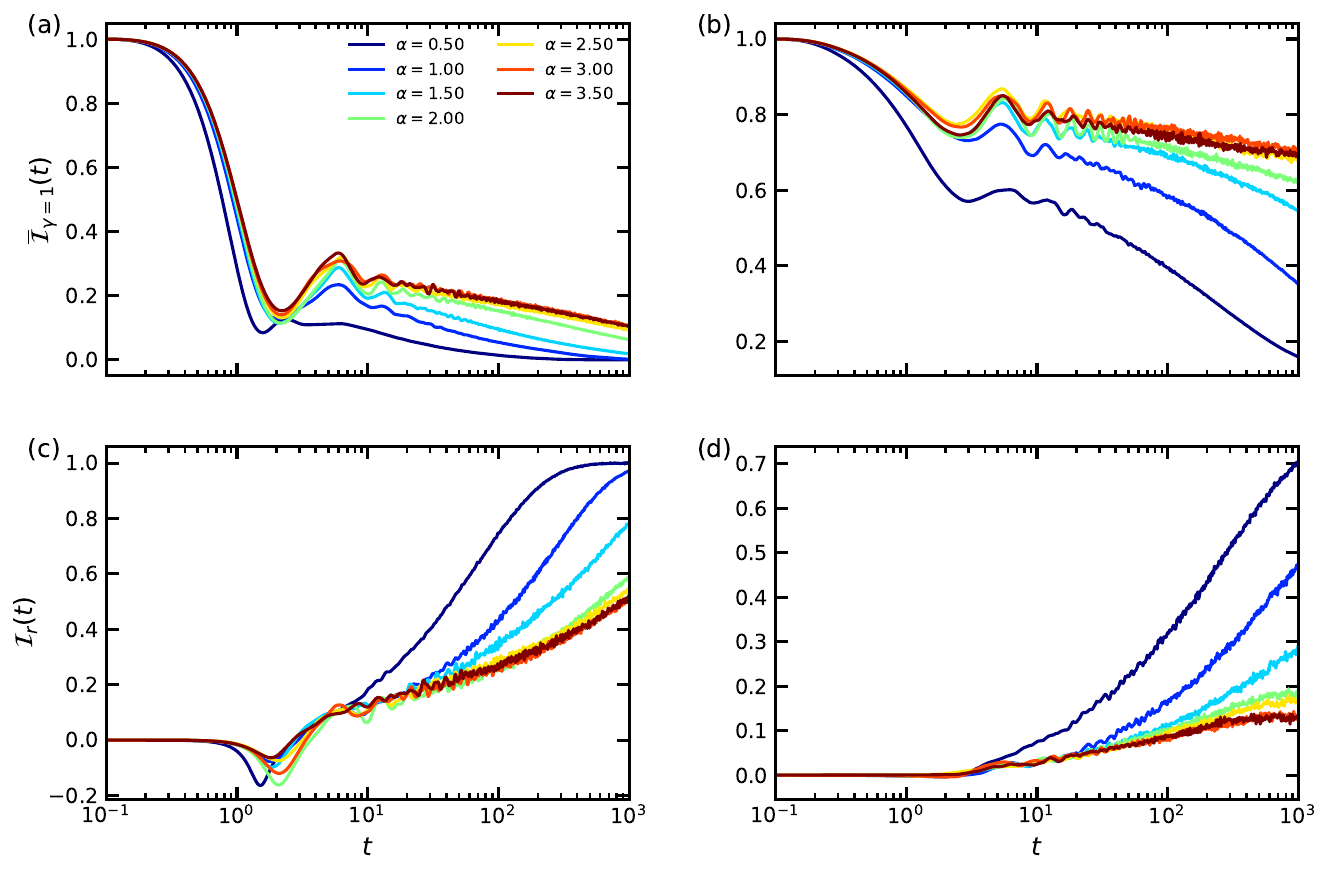}
\caption{Magnetization imbalance dynamics for a fixed system size $L=16$ coupled to a thermal bath. (a) and (b) display the raw magnetization imbalance $\overline{\mathcal{I}}_{\gamma=1}(t)$ under weak disorder ($W=2$) and strong disorder ($W=15$), respectively. (c) and (d) show the corresponding relative magnetization imbalance $\mathcal{I}_r(t)$ for the same disorder strengths. Colors represent the interaction exponents, ranging from $\alpha=0.5$ (dark blue) to $\alpha=3.5$ (dark red). (a) For all $\alpha$, the system rapidly reaches thermal equilibrium, with the relaxation rate decreasing monotonically as $\alpha$ increases. (b) A pronounced separation of timescales is observed: smaller $\alpha$ results in rapid decay, while larger $\alpha$ (e.g., $\alpha=3.5$) decays more slowly, indicating that the thermalization time increases exponentially with the interaction range. (c) For smaller $\alpha$, $\mathcal{I}_r(t)$ rapidly approaches 1, signifying complete thermalization; the rate of approach decreases monotonically with increasing $\alpha$. (d) A significant divergence in the dynamics is observed: for smaller $\alpha$ (blue curves), $\mathcal{I}_r(t)$ grows rapidly, indicating efficient avalanche propagation. Conversely, for larger $\alpha$ (red curves), $\mathcal{I}_r(t)$ maintains a slow growth and close to zero., indicating that avalanches are effectively inhibited and the system retains memory of its initial state.}
\label{Fig6}
\end{figure}

To further analyze the influence of $\alpha$ on the propagation velocity of the bath-induced thermalization front, we fix the system size at $L=16$ and examine the evolution of the magnetization imbalance $\overline{\mathcal{I}}_{\gamma=1}(t)$ and the relative magnetization imbalance $\mathcal{I}_r(t)$ for various values of $\alpha$, as shown in Fig.~\ref{Fig6}. Figures~\ref{Fig6}(a) and~\ref{Fig6}(b) compare the behavior of $\overline{\mathcal{I}}_{\gamma=1}(t)$ in the weak disorder ($W=2$) and strong disorder ($W=15$) regimes, respectively. In the weak disorder regime (Fig.~\ref{Fig6}(a)), the system generally exhibits ergodic behavior. Although the imbalance eventually decays to zero for all values of $\alpha$, a clear monotonic trend is observed: the relaxation rate decreases as $\alpha$ increases. Specifically, for smaller $\alpha$ values, the decay is nearly instantaneous and independent of system size; for larger $\alpha$ values, the process is slightly slower but qualitatively similar. Overall, the qualitative behavior remains consistent—the thermal bath successfully drives the system to thermal equilibrium within a relatively short timescale ($t \sim 10^2$), regardless of the interaction range, which is consistent with ETH predictions.

In the strong-disorder regime ($W=15.0$, Fig.~\ref{Fig6}(b)), the physical behavior becomes more transparent. In this parameter region, the system has already crossed the phase boundary between the ETH and MBL phases of the corresponding static model. As a result, the interaction exponent $\alpha$ provides an efficient tuning knob to strongly modify the dynamical response, and thus serves as a key parameter controlling the stability of the system. For small $\alpha$ (e.g., $\alpha=0.5$, dark-blue curve), the system remains ergodic and the imbalance decays rapidly and substantially. Even at such a high disorder strength, the enhanced connectivity induced by long-range interactions enables the bath-triggered “avalanche” to propagate efficiently. Upon increasing $\alpha$, the dynamics slows down markedly and thermalization is strongly suppressed. In particular, for the shortest-range case considered here ($\alpha=3.5$, dark-red curve), the imbalance exhibits remarkable stability and shows only negligible decay throughout the entire simulation time window $t\in[10^{-1},10^{3}]$. Within this regime, the thermalization time exceeds our observation limit, allowing us to conclude that the MBL phase can effectively withstand bath-induced avalanches. The pronounced differences among different $\alpha$ further indicate that the thermalization timescale depends exponentially on the interaction range, corroborating that increasing $\alpha$ can drive the system from an avalanche-prone fragile regime into a robust MBL phase where thermalization dynamics is strongly hindered.

Similarly, in Figs.~\ref{Fig6}(c) and~\ref{Fig6}(d), we analyze the evolution of the relative magnetization imbalance $\mathcal{I}_r(t)$ for different $\alpha$ in the weak disorder ($W=2$) and strong disorder ($W=15$) regimes, respectively. $\mathcal{I}_r(t)$ serves as a normalized measure of the "damage" caused by avalanches, where $\mathcal{I}_r(t) \to 0$ signifies a completely insulating system, while $\mathcal{I}_r(t) \to 1$ indicates full thermalization. In the ergodic phase ($W=2$, Fig.~\ref{Fig6}(c)), the behavior of the relative imbalance aligns with the expectations for a thermalizing system. Regardless of the interaction exponent $\alpha$, all curves eventually saturate near $\mathcal{I}_r \approx 1$. However, the interaction range dictates the timescale of this process: long-range interactions (small $\alpha$, dark blue lines) facilitate rapid energy exchange with the environment, leading to a sharp rise in $\mathcal{I}_r(t)$. As $\alpha$ increases (dark red lines), the coupling between bulk spins and environment-induced excitations weakens, slowing down the thermalization process; nevertheless, the ultimate ergodic outcome remains inevitable.

In the strong-disorder regime ($W=15.0$, Fig.~\ref{Fig6}(d)), the dynamics exhibits qualitatively different behavior, where $\mathcal{I}_r(t)$ serves as a sensitive probe of the stability of the MBL phase against bath-induced avalanches. For long-range interactions (dark-blue curve), $\mathcal{I}_r(t)$ increases rapidly in time and reaches appreciable values within the simulated time window. This indicates that the avalanche process triggered by the bath can propagate efficiently, thereby driving the chain progressively toward thermalization; it also implies that when the interaction decays slowly, avalanche spreading remains highly effective. In contrast, as $\alpha$ is increased toward the short-range limit (dark-red/red curves), the growth of $\mathcal{I}_r(t)$ is strongly suppressed, and the curves flatten and approach zero, suggesting that the bath-induced “thermal bubbles” cannot effectively proliferate through the rigid localized bulk. The sharp contrast between the rapidly rising blue curve and the strongly suppressed red curves in Fig.~\ref{Fig6}(d) provides a direct manifestation of our main conclusion: upon tuning the interaction range, the MBL phase undergoes a crossover from avalanche instability to stability, and for sufficiently large $\alpha$ the avalanche propagation speed decreases exponentially.

\begin{figure*}[t!]
\centering
\includegraphics[width=1\textwidth,keepaspectratio]{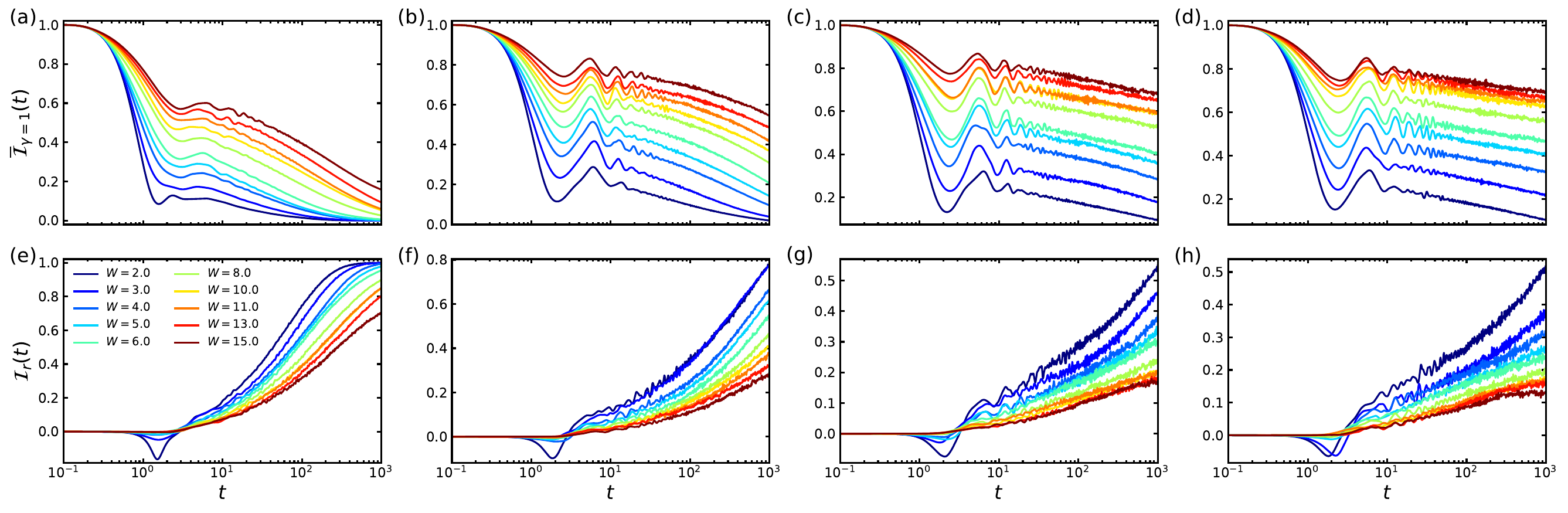}
\caption{Dynamical evolution of the magnetization imbalance under the coupling to a thermal bath for a fixed system size $L=16$ and different disorder strengths $W$. The top row (a)–(d) displays the absolute imbalance $\overline{\mathcal{I}}_{\gamma=1}(t)$, while the bottom row (e)–(h) shows the relative imbalance $\mathcal{I}_r(t)$. Each column corresponds to an interaction exponent $\alpha = 0.5, 1.5, 2.5, 3.5$ from left to right. The curve colors represent the disorder strength $W$, ranging from the ergodic regime ($W=2$, dark blue) to the deep MBL regime ($W=15$, dark red). A distinct crossover behavior is observed: for long-range interactions (left columns), increasing $W$ only slightly retards thermalization, whereas for short-range interactions (right columns), strong disorder effectively suppresses avalanches, leading to a significant divergence between ergodic and localized trajectories.}
\label{Fig7}
\end{figure*}

To obtain a comprehensive characterization of the interplay between disorder strength and interaction range in suppressing quantum avalanches, we analyze the time evolution of the magnetization dynamics for a fixed system size $L=16$ while varying the disorder strength from $W=2$ to $W=15$. Figure~\ref{Fig7} illustrates this dependence, with the top row displaying the absolute imbalance $\overline{\mathcal{I}}_{\gamma=1}(t)$ and the bottom row showing the relative imbalance $\mathcal{I}_r(t)$. A distinct shift in dynamical behavior occurs as the interaction transitions from long-range (left columns) to short-range (right columns). In the strong long-range interaction regime ($\alpha=0.5$), increasing disorder provides only weak protection against bath-induced thermalization; even for the maximum disorder value ($W=15$, red curve), the relative imbalance $\mathcal{I}_r(t)$ rises significantly, indicating that the avalanche front permeates the entire system despite the presence of a strong random potential. Conversely, as the interaction range shortens ($\alpha \ge 2.5$), the suppressive effect of disorder on the dynamics becomes increasingly pronounced. This is manifested by the prominent "fanning out" of the curves in the rightmost panels: weak disorder (blue curves) still allows for rapid thermalization, whereas strong disorder (red curves) effectively inhibits the growth of $\mathcal{I}_r(t)$, keeping it near zero throughout the simulation window. This contrast underscores that the stability of the MBL phase is not solely determined by disorder strength but is significantly modulated by the interaction range, with short-range interactions being a prerequisite for disorder to effectively block avalanche.

\begin{figure*}[t!]
\centering
\includegraphics[width=1\textwidth,keepaspectratio]{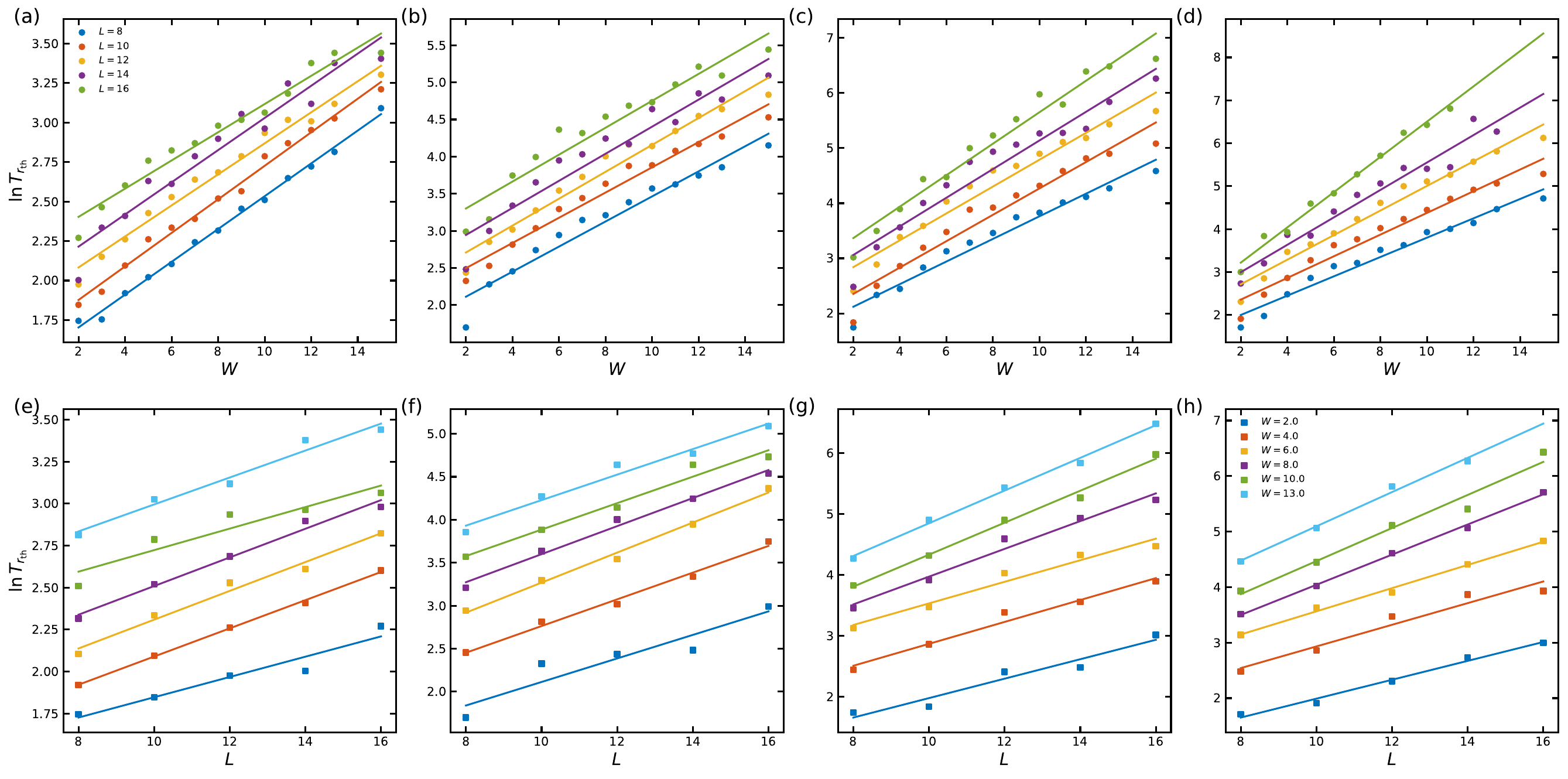}
\caption{Scaling analysis of the characteristic thermalization time $\ln T_{r_{\mathrm{th}}}$, showing its dependence on the system size $L$ and the disorder strength $W$. Columns from left to right correspond to interaction exponents $\alpha = 0.5,1.5,2.5,$ and $3.5$. The upper row (a)–(d) shows $\ln T_{r_{\mathrm{th}}}$ as a function of $W$ for different system sizes $L\in{8,10,12,14,16}$, while the lower row (e)–(h) shows $\ln T_{r_{\mathrm{th}}}$ as a function of $L$ for different disorder strengths $W$. The semilogarithmic representation highlights the scaling form: over broad parameter ranges, $\ln T_{r_{\mathrm{th}}}$ exhibits an approximately linear dependence on both $W$ and $L$, with the corresponding slopes increasing with $\alpha$.}
\label{Fig8}
\end{figure*}

To extract the relevant timescale governing the avalanche dynamics and to quantitatively assess the stability of the MBL phase, we introduce a characteristic thermalization time $T_{r_{\mathrm{th}}}$, defined as the earliest time at which the relative imbalance $\mathcal{I}_r(t)$ first exceeds a prescribed threshold $r{\mathrm{th}}$. The choice of $r_{\mathrm{th}}$ must balance the representativeness of the thermalization criterion and numerical accessibility: if $r_{\mathrm{th}}$ is too small, it may be dominated by early-time fluctuations and thus fail to capture the onset of thermalization; conversely, if $r_{\mathrm{th}}$ is too large, the corresponding crossing time grows exponentially for large $\alpha$ and strong disorder $W$, exceeding the time window accessible in our calculations and leading to substantially increased statistical uncertainty. Therefore, following Ref.~\cite{Scocco2024}, we fix $r_{\mathrm{th}}=0.17$ throughout this work.

The dependence of $T_{r_{\mathrm{th}}}$ on disorder strength $W$ and system size $L$ is summarized in Fig.~\ref{Fig8}. We note that for the fixed threshold $r_{\mathrm{th}}=0.17$, in some parameter regimes the relative imbalance $\mathcal{I}_r(t)$ does not reach the threshold within the accessible simulation window $t\le 10^3$. This occurs most prominently in the strongly disordered and short-range regime, e.g., for $W=15$ at large $\alpha$ [Figs.~\ref{Fig8}(d) and~\ref{Fig8}(h)], where the growth of $\mathcal{I}_r(t)$ is strongly suppressed. In such cases, the corresponding $T_{r_{\mathrm{th}}}$ should be interpreted as exceeding our numerical time horizon, i.e., $T_{r_{\mathrm{th}}} > 10^3$, providing a lower bound on the characteristic thermalization time. This behavior further reinforces the drastic slowdown of bath-induced thermalization in the avalanche-inhibited localized regime.

The first row of Fig.~\ref{Fig8} shows the dependence of the characteristic thermalization time $T_{r_{\mathrm{th}}}$ on the disorder strength $W$. Overall, $\ln T_{r_{\mathrm{th}}}$ increases approximately linearly with $W$, $\ln T_{r_{\mathrm{th}}}\propto W$, indicating that stronger disorder generically slows down the dynamics across all interaction ranges. A more stringent test of the stability of the MBL phase is provided by the system-size scaling of $T_{r_{\mathrm{th}}}$, shown in the second row of Fig.~\ref{Fig8}. In the semilogarithmic representation, $\ln T_{r_{\mathrm{th}}}$ exhibits an approximately linear dependence on $L$ over broad parameter regimes, $\ln T_{r_{\mathrm{th}}}\propto L$, which corresponds to an exponential scaling of the thermalization time with system size. Notably, as the interactions are tuned from the long-range regime ($\alpha=0.5$) toward the short-range limit ($\alpha=3.5$), the slopes of the curves progressively increase, demonstrating that the sensitivity of the thermalization time to both disorder strength and system size depends strongly on the interaction exponent $\alpha$. In the strongly long-range case ($\alpha=0.5$, Fig.~\ref{Fig8}(a) and Fig.~\ref{Fig8}(e)), the slopes of $\ln T_{r_{\mathrm{th}}}$ as a function of both $W$ and $L$ remain relatively small, implying a weak dependence of the thermalization time on disorder and system size. This behavior is consistent with delocalization/ergodicity, where bath-induced thermalization efficiently penetrates the entire chain with only a weak dependence on chain length. By contrast, in the short-range regime (e.g., $\alpha=3.5$, Fig.~\ref{Fig8}(d) andFig.~\ref{Fig8}(h)), $\ln T_{r_{\mathrm{th}}}$ displays a pronounced approximately linear increase with both $W$ and $L$, implying an exponential growth of the thermalization time with disorder strength and system size. Such exponential scaling constitutes a key signature of a stable MBL phase in the presence of an avalanche seed: it indicates that bath-induced thermal “bubbles” cannot proliferate through the localized bulk, so that the time required to thermalize the entire system becomes exceedingly long in the thermodynamic limit. These results therefore demonstrate that, for sufficiently large $\alpha$, the localized phase is robust against quantum avalanches.

\begin{figure*}[t!]
\centering
\includegraphics[width=1\textwidth,keepaspectratio]{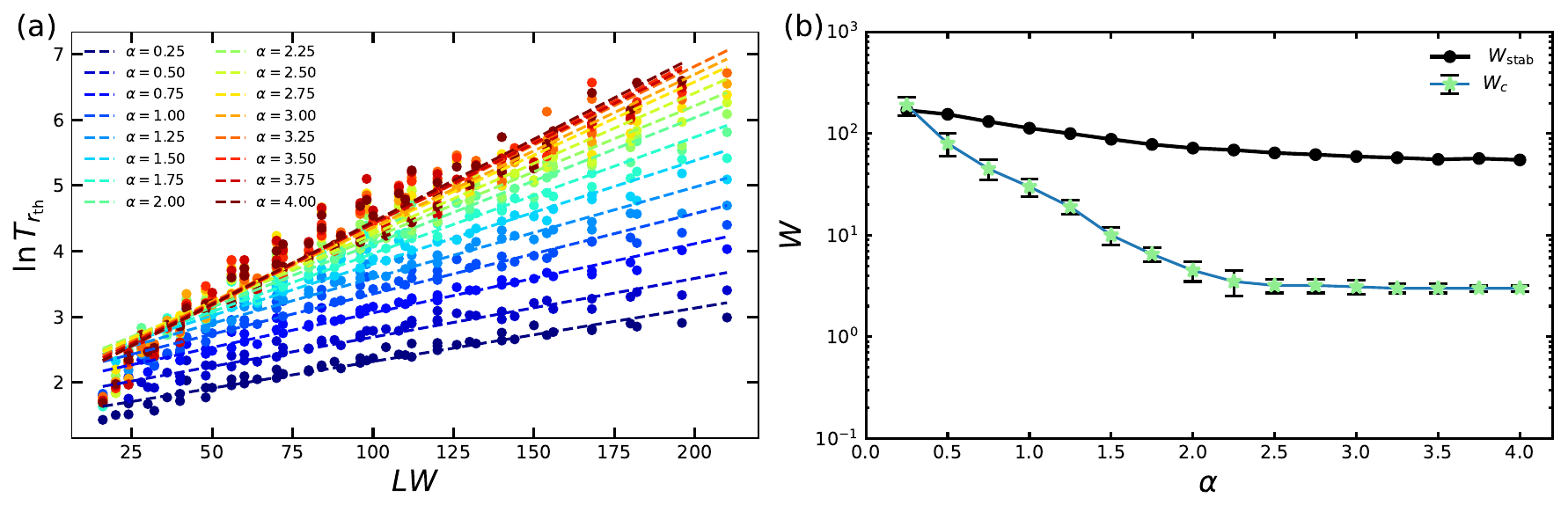}
\caption{Unified scaling analysis of the characteristic thermalization time $T_{r_{\text{th}}}$. (a) $T_{r_{\text{th}}}$ (defined at a relative imbalance threshold of $0.17$) is plotted as a function of the composite scaling variable $LW$. Data points aggregate results across various system sizes and disorder strengths, color-coded by the interaction exponent $\alpha \in [0.25, 4.00]$ (dark blue to dark red). Dashed lines represent linear fits of the form $\ln{T_{r_{\text{th}}}} \propto \kappa(\alpha) LW$. The slope $\kappa(\alpha)$ increases monotonically with $\alpha$, indicating that for short-range interactions (larger $\alpha$), the effective thermalization barrier grows rapidly with the system's insulating parameter $LW$, whereas this growth is significantly suppressed for long-range interactions (smaller $\alpha$). (b) Comparison between the minimum disorder strength required for stability against the thermal bath, $W_{\text{stab}}$, and the critical disorder strength $W_c$ calculated at infinite temperature.}
\label{Fig9}
\end{figure*}

To provide a comprehensive description of avalanche dynamics and unify the influences of system size $L$, disorder strength $W$, and interaction range $\alpha$, we investigate the scaling behavior of the thermalization time with respect to the composite variable $LW$. As shown in Fig.~\ref{Fig9}(a), when plotting the logarithm of the thermalization time $\ln T_{r_{\text{th}}}$ against $LW$, the data points for a fixed $\alpha$ collapse well onto linear trajectories. This linearity provides strong evidence for a unified scaling law of the form:
$$
T_{r_{\text{th}}} \sim \exp(\kappa(\alpha) LW).
$$

To evaluate the impact of different $\alpha$ values on the avalanche velocity, we perform linear fits using the functional form $\ln T_{r_{\text{th}}} = \kappa(\alpha) LW + \beta$, where each $\alpha$ corresponds to a specific slope $\kappa(\alpha)$, and $\beta$ is a constant intercept. These results suggest that the localized bulk acts as an effective “barrier” (or insulating layer) against avalanche propagation, whose blocking strength increases with the product $LW$ of the system size and the disorder strength. In other words, increasing either $W$ or $L$ substantially enhances the timescale required for bath-induced avalanches to penetrate through the localized region. More importantly, the interaction exponent $\alpha$ controls the effective strength of this barrier through the slope $\kappa(\alpha)$. In the long-range regime (dark-blue fit), the small value of $\kappa(\alpha)$ indicates that increasing $W$ or $L$ yields only a limited delay of thermalization, since long-range couplings enhance the connectivity and partially circumvent the blocking effect imposed by the random potential. By contrast, upon approaching the short-range limit (red fit), $\kappa(\alpha)$ increases markedly and eventually saturates, quantitatively demonstrating that short-range interactions strongly enhance the suppression of avalanche spreading by the localized bulk. For sufficiently large $\alpha$, the timescale for an avalanche to traverse the system grows rapidly and exponentially with $LW$, confirming that the short-range limit is crucial for maximizing the stability of the system against quantum avalanches.

For a finite-size system of length $L$, Morningstar \textit{et al.} proposed an avalanche-stability criterion stating that the MBL phase can be regarded as stable against avalanche effects if the slowest thermalization time $t_s$ grows faster than $4^L$ with increasing system size, implying that thermalization is effectively delayed in the thermodynamic limit~\cite{Sels2022,Scocco2024,Morningstar2022}. Motivated by this criterion, we further assume that in the limit $L\to\infty$ there exists a disorder threshold $W_{\mathrm{stab}}$ such that the system remains robust against bath-induced avalanches for $W>W_{\mathrm{stab}}$. For $W<W_{\mathrm{stab}}$, the criterion no longer guarantees stability: the system may still remain localized, but it may also ultimately undergo avalanche-induced delocalization and thermalize on very long time scales. Therefore, $W_{\mathrm{stab}}$ should be interpreted as a conservative stability threshold (a sufficient bound), rather than a sharp phase boundary.

To quantitatively estimate $W_{\mathrm{stab}}$, we employ the scaling form for the thermalization time obtained above
\begin{equation}
\ln(t_e)=\kappa(\alpha)\,LW+\beta,
\label{eq:te_scaling}
\end{equation}
where $t_e$ denotes the characteristic time required for the relative imbalance to cross the threshold $r_{\mathrm{th}}$ for the first time. Since the slowest thermalization time satisfies the lower-bound relation $t_s\gtrsim t_e$, requiring $t_e$ to grow faster than $4^L$ ensures that $t_s$ also satisfies the Morningstar criterion, thereby providing a conservative stability condition. Imposing $t_e>4^L$ (equivalently, $\ln t_e>L\ln 4$) and substituting into Eq.~\eqref{eq:te_scaling} yields the minimal disorder strength required for stability at finite $L$
\begin{equation}
W_{\mathrm{stab}}(L)=\frac{2\ln 2}{\kappa(\alpha)}-\frac{\beta}{\kappa(\alpha)L}.
\label{eq:Wstab_L}
\end{equation}

Furthermore, assuming that the $LW$ scaling behavior observed in Fig.~9(a) holds over a sufficiently broad parameter regime, we obtain in the thermodynamic limit $L\to\infty$ the disorder threshold
\begin{equation}
W_{\mathrm{stab}}=\frac{2\ln 2}{\kappa(\alpha)}.
\label{eq:Wstab_inf}
\end{equation}

Based on Eq.~\eqref{eq:Wstab_inf}, we extract $\kappa(\alpha)$ from linear fits and obtain the $\alpha$ dependence of $W_{\mathrm{stab}}$, as shown in Fig.~9(b). For comparison, Fig.~9(b) also displays the infinite-temperature MBL critical line $W_c(\alpha)$ for the corresponding isolated system. It is important to emphasize that $W_c$ characterizes the ETH--MBL critical behavior in the static (closed) system, whereas $W_{\mathrm{stab}}$ quantifies the stability threshold of the MBL phase against bath-induced thermalization in the presence of an avalanche seed. Accordingly, $W_{\mathrm{stab}}$ generally lies deeper in the disordered regime and provides a more stringent condition for localization. Specifically, for $W>W_{\mathrm{stab}}$ we can conclude with high confidence that the localized phase remains stable against avalanches. In contrast, in the intermediate regime $W_c<W<W_{\mathrm{stab}}$, the system may display MBL-like behavior at finite sizes and accessible time scales, while still being susceptible to avalanche-induced thermalization in the thermodynamic limit. This regime can be interpreted as an ``asymptotically localized'' (asymptotic MBL or quasi-MBL) dynamical window.

From a physical perspective, such asymptotic localization can be understood in terms of many-body resonances. For sufficiently strong long-range interactions (smaller $\alpha$), the coupling between rare thermal regions and the surrounding localized bulk is enhanced, making rare regions more effective in triggering many-body resonances and thereby facilitating avalanche propagation. Although in finite systems most eigenstates may remain nonresonant and exhibit MBL-like dynamical signatures, increasing the system size enhances both the probability and the spatial influence of such rare resonant regions, ultimately leading to thermalization. As $\alpha$ is further reduced, the asymptotic-MBL window correspondingly shrinks and eventually disappears, reflecting the pronounced enhancement of avalanche spreading by long-range interactions.

\section{Conclusions}\label{sec_conclusion}

In this work, we investigated the robustness of MBL against bath-induced quantum avalanches in one-dimensional Heisenberg chains with power-law interactions ($V\sim r^{-\alpha}$). Going beyond isolated-system diagnostics, we employed a hybrid framework that combines ED (to locate the ETH--MBL transition boundary in the $(\alpha,W)$ plane) with Lindblad open-system simulations, where one end of the chain is coupled to an infinite-temperature bath. This setup provides a minimal and well-controlled dynamical protocol to probe how an ergodic inclusion triggers thermalization and whether the localized bulk can inhibit the resulting avalanche.

Our central finding is a quantitative dynamical stability criterion for long-range MBL. By analyzing the bath-induced growth of the relative imbalance $\mathcal{I}_r(t)$ and defining a characteristic thermalization time $T_{r_{\mathrm{th}}}$ via a fixed threshold crossing, we uncover a unified scaling law $
T_{r_{\mathrm{th}}}\sim \exp\!\big[\kappa(\alpha)\,LW\big],$ demonstrating that avalanche-driven thermalization becomes exponentially slow in both system size and disorder strength, with a rate controlled by the interaction range through $\kappa(\alpha)$. Leveraging this scaling form together with the avalanche-stability criterion of Ref.~\cite{Morningstar2022}, we extract an interaction-dependent threshold disorder $W_{\mathrm{stab}}(\alpha)$, which provides a conservative bound for stability in the thermodynamic limit: for $W>W_{\mathrm{stab}}(\alpha)$, bath-induced avalanches are dynamically suppressed and the localized phase remains robust, whereas for smaller disorder the fate of localization cannot be guaranteed. Importantly, $W_{\mathrm{stab}}(\alpha)$ increases sharply as interactions become longer ranged, establishing that shortening the interaction range is crucial for stabilizing MBL against avalanches. These results deliver a controlled bridge between static phase boundaries and real-time avalanche dynamics, and directly motivate experimental tests in AMO platforms with tunable $\alpha$ and engineered dissipation.

\appendix

\section{Stability Analysis of Fitting Parameters}\label{sec:appendix_stability}

\begin{figure}[t!]
\centering
\includegraphics[width=0.45\textwidth,keepaspectratio]{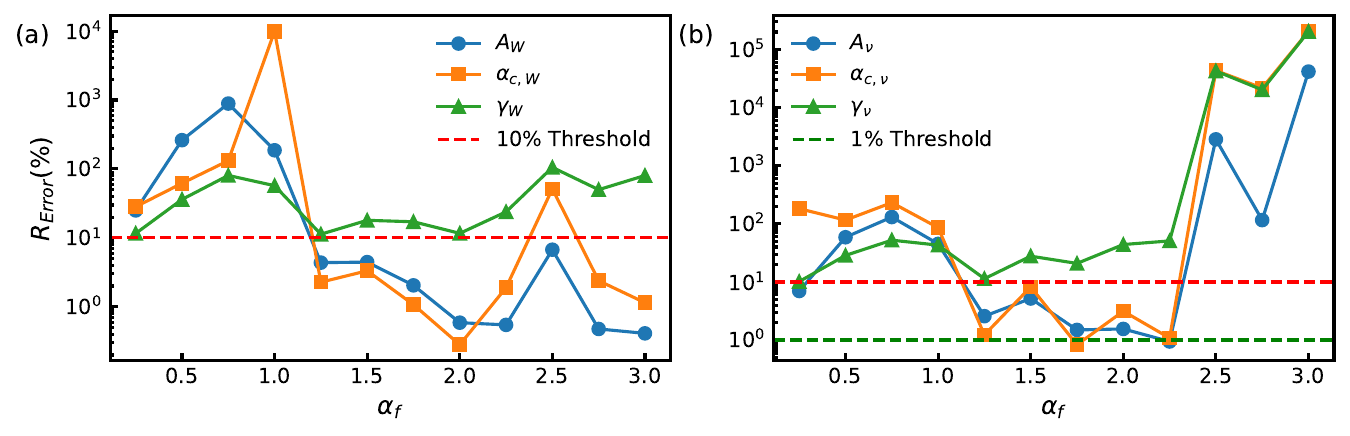}
\caption{Parameter stability analysis employed to determine the optimal lower-bound cutoff $\alpha_f$ for the power-law scaling fits. The plots display the relative error ($R_{\text{Error}}$, in percentage) of the fitting coefficients as a function of the cutoff $\alpha_f$. (a) Analysis for the critical disorder $W_c$. The relative errors for $A_W$ (blue circles), $\alpha_{c,W}$ (orange squares), and $\gamma_W$ (green triangles) decrease significantly as the cutoff excludes the long-range regime, showing a clear stability minimum at $\alpha_f = 2$ where all errors fall below the $10\%$ threshold (red dashed line). (b) Analysis for the critical exponent $\nu$. A robust stability window is observed for $\alpha_f \in [1.25, 2.25]$, where errors approach the high-precision $1\%$ threshold (green dashed line). For $\alpha_f > 2.5$, the errors diverge sharply due to the insufficient number of data points available for the fit.}
\label{Fig10}
\end{figure}

In this Appendix, we provide the technical details supporting the determination of the cutoff value $\alpha_f$ used in the power-law fits presented in Sec.~\ref{sec_isolated}. As the interaction exponent $\alpha$ decreases towards the critical value $\alpha_c$, finite-size effects and the crossover from power-law to exponential scaling can distort the true divergence. Consequently, defining an appropriate data window $\alpha > \alpha_f$ is essential to extract universal critical parameters.

Figure~\ref{Fig10} illustrates the relative error of the fitting parameters ($A_\eta, \alpha_{c,\eta}, \gamma_\eta$) as a function of the cutoff $\alpha_f$ for both the critical disorder $W_c$ [Fig.~\ref{Fig10}(a)] and the critical exponent $\nu$ [Fig.~\ref{Fig10}(b)]. The relative error is defined as the ratio of the standard deviation to the estimated value of the parameter as returned by the nonlinear least-squares algorithm.

For the critical disorder $W_c$, shown in Fig.~\ref{Fig10}(a), we observe that for small cutoffs ($\alpha_f < 1.2$), the errors are substantial—exceeding $100\%$ for the critical exponent $\alpha_{c,W}$—indicating that the inclusion of data from the long-range exponential regime contaminates the power-law scaling ansatz. As $\alpha_f$ increases, the errors decrease sharply. We identify a distinct region of stability around $\alpha_f \approx 2$, where the relative errors for all three parameters drop well below the $10\%$ empirical threshold (red dashed line). Furthermore, $\alpha_f = 2$ represents the global minimum for the error in $\alpha_{c,W}$, justifying its selection for our final analysis.

For the critical exponent $\nu$, displayed in Fig.~\ref{Fig10}(b), the error landscape exhibits a clear "valley" of stability. In the intermediate range $\alpha_f \in [1.25, 2.25]$, the fitting procedure is highly robust, with errors approaching the stringent $1\%$ threshold (green dashed line). Conversely, if the cutoff is chosen too large ($\alpha_f > 2.5$), the number of available data points becomes insufficient to constrain the three-parameter model, leading to a divergence in the fitting uncertainties. We selected $\alpha_f = 1.75$ as the optimal cutoff, as it lies centrally within the stable region and minimizes the combined uncertainty for all fitting parameters.

\section*{Acknowledgments}

This work is supported by the National Natural Science Foundation of China (11975175).

\section*{DATA AVAILABILITY STATEMENT}

All raw data corresponding to the findings in this paper are available from the authors upon
reasonable request.

\section*{Author declarations}

The authors have no conflicts to disclose.

\bibliography{references_main.bib}

\end{document}